\documentclass[pra,twocolumn,showpacs,letterpaper,showpacs,superscriptaddress]{revtex4}
\usepackage{graphicx,amsmath,amssymb,amsfonts,latexsym,color,dcolumn,bm,epsfig,subfigure}
\def\bbm[#1]{\mbox{\boldmath $#1$}}
\def\TE{\mathrm{TE}}
\def\TM{\mathrm{TM}}

\def\br{\mathbf{r}}
\def\bk{\mathbf{k}}

\begin{document}

\title{Dispersive interactions between atoms and non planar surfaces}

\author{Riccardo Messina}
\affiliation{Laboratoire Kastler Brossel, case 74,
CNRS, ENS, UPMC, Campus Jussieu, F-75252 Paris Cedex 05, France}
\affiliation{Dipartimento di Scienze Fisiche e Astronomiche
dell'Universita' degli Studi di Palermo and CNSIM, Via Archirafi 36,
I-90123 Palermo, Italy}

\author{Diego A. R. Dalvit}
\affiliation{Theoretical Division, Los Alamos National Laboratory, Los Alamos, NM 87545, USA}

\author{Paulo A. Maia Neto}
\affiliation{Instituto de F\'{\i}sica, UFRJ, CP 68528, Rio de Janeiro,  RJ, 21941-972, Brazil}

\author{Astrid Lambrecht}
\affiliation{Laboratoire Kastler Brossel, case 74,
CNRS, ENS, UPMC, Campus Jussieu, F-75252 Paris Cedex 05, France}

\author{Serge Reynaud}
\affiliation{Laboratoire Kastler Brossel, case 74,
CNRS, ENS, UPMC, Campus Jussieu, F-75252 Paris Cedex 05, France}

\date{\today}

\begin{abstract}
We calculate the dispersive force between a ground state atom and a non planar surface.
We present explicit results for a corrugated surface, derived from the scattering
approach at first order in the corrugation amplitude.
A variety of analytical results are derived in different limiting cases,
including the van der Waals and Casimir-Polder regimes.
We compute numerically the exact first-order dispersive potential for
arbitrary separation distances and corrugation wavelengths,
for a Rubidium atom on top of a silicon or gold corrugated surface.
We discuss in detail the inadequacy of the proximity force approximation,
and present a simple but adequate approximation for computing the potential.
\end{abstract}


\maketitle


\section{Introduction}

Quantum and thermal fluctuations of the electromagnetic field in the presence of material
boundaries generate fluctuating spatial gradients of the field intensity and fluctuating
induced electric dipoles in ground-state atoms close to those surfaces, resulting in an
atom-surface interaction via the optical dipole force. These dispersive forces have
considerable importance in fundamental physics, including deflection of atomic beams close
to surfaces \cite{deflection}, classical \cite{Aspect1996} and quantum reflection of cold
atoms \cite{Shimizu2001,DeKieviet2003} and Bose-Einstein condensates (BECs) \cite{QRbec}
from surfaces, dipole oscillations of BECs above dielectric surfaces \cite{Cornell05,Cornell07},
and in possible future applications of single-atom manipulation in atom chips \cite{atomchip,CAMS05}.
At the limit of large separation distances (Casimir-Polder limit \cite{CasimirPolder}),
the interactions show universal features since they depend only
on the zero-frequency atom and surface optical responses.

Dispersive forces can be tailored in different ways, either by engineering the optical properties of
the surfaces \cite{Buhmann07,Rosa08,Pitaevskii08}, by biasing the temperatures of the surfaces and
their thermal environment \cite{Cornell07,Antezza05,Buchleitner08}, or by suitably changing their geometry.
Given the complexity of these quantum forces, which arise from fluctuations at all
frequency and length scales, calculations beyond planar geometries are exceedingly involved.
Until recently, approximate methods have been used to deal with non-planar setups, including the proximity
force approximation (PFA) \cite{Derjaguin57} and the pairwise summation approach (PWS) \cite{Galina00}.
However, these approximate methods drastically fail when the surface is not sufficiently smooth on the
length-scale of the atom-surface separation, since they do not properly take into account the non-additivity
of dispersion forces \cite{JPA}. These large deviations from PFA or PWS could be probed by using present-day
technology, for example with a Bose-Einstein condensate above a microfabricated corrugated surface \cite{PRL}.
Similar deviations have been recently measured for the Casimir force between a metallic sphere and a
nanostructured silicon rectangular grating \cite{Chan08,Lambrecht08}.

A number of methods are available to compute the dispersive atom-surface forces
(see for example \cite{CasimirPolder,Barton,WylieSipe,Mostepanenko,Buhman05}).
These methods allow the exact computation of the force for very simple geometries, such as
an atom above a single (possibly multilayered) plane, sphere or cylinder, or an atom inside
a plane Fabry-Perot cavity. Toy models with scalar fields with ideal (Dirichlet) boundary conditions
have also been considered to compute the ``scalar Casimir-Polder" force for a small sphere above a
uni-axial corrugated surface \cite{Gies08}.

Similar problems have been discussed for the Casimir interaction between two macroscopic bodies
(see for example \cite{Lifshitz56,DLP,Kats77}),
and the scattering approach has been shown to be a very efficient tool for analyzing
the case of non-trivial geometries \cite{Lambrecht06}.
{}For parallel plates, the Casimir force may be written in terms of the reflection coefficients
seen from the region in-between the plates \cite{JaekelReynaud91,Genet2003}.
A similar expression also holds for non-planar surfaces, with the specular reflection coefficients
replaced by more general reflection operators that describe non-specular diffraction by the
surfaces. This allows for the computation of the Casimir force between
rough \cite{MaiaNeto05} or corrugated \cite{Rodrigues06} surfaces. Similar methods have been
employed to compute the force between a plane and a sphere \cite{Bulgac06,Bordag06,Emig08,MaiaNeto08}
and between two spheres \cite{Emig07,Kenneth08}.

In this paper we develop the scattering approach for studying the dispersive interaction between
ground-state atoms and arbitrarily shaped surfaces characterized by frequency-dependent reflection operators.
This approach provides an exact analytical expression for the two-body interaction energy, which can be written
in terms of the scattering matrices defined for each individual, one-body, scatterer.
Although this is a simpler problem than solving for the many-body Green function, it is still very difficult
to find the exact scattering matrices for the problem of a
electromagnetic field impinging on a given non-planar geometry.
In order to illustrate the method, we write the two-body interaction energy in a perturbative expansion in terms
of the deviation of the surface profile from the planar case \cite{PRL}. Apart from this perturbative approximation,
the approach is valid for any value of the corrugation wavelength and includes the full electromagnetic fluctuations
and the real material properties of the boundaries. We  compute the perturbative dispersive interaction potential
to first order in the surface profile both for ideal and real materials, and discuss the prospects of measuring
non-trivial (beyond PFA) geometrical effects with a uni-axial corrugated surface interacting with a Bose-Einstein
condensate used as a vacuum field sensor.

The paper is organized in the following way. In Sec. 2, we develop the scattering formalism for the
atom-surface interaction. The resulting general formula is then applied to calculate the potential to first order
in the surface corrugation amplitude in Sec. 3.
We compare our results with PFA in Sec. 4 and discuss experimental prospects  in Sec. 5.
Concluding remarks are presented in Sec. 6.


\section{Atom-surface interaction for non-planar geometries}

We consider a ground-state atom at position ${\bf R}_A=(\br_A,z_A)$  on top of a non-planar surface, as shown by Fig.~1.
The surface profile is defined by the function $h(x,y)$ representing the local height with respect to the plane $z=0.$
The atom is in free space, so that $z_A>h(\br_A).$
The fluctuating electromagnetic field propagating
from the surface towards the atom is written in the mixed Fourier representation
(frequency $\omega$, two-dimensional wave vector ${\bf k}$)
\begin{equation}
\label{representation}
{\bf E}^{\uparrow}({\bf k}, z, \omega) =\sum_{p=\TE,\TM}{E}_p^{\uparrow}({\bf k}, \omega)\mbox{\boldmath\({\hat \epsilon}\)}
^+_p({\bf k},\omega)e^{i k_z z},
\end{equation}
where $p$ is the polarization index and $k_z$ the longitudinal wavevector (sgn denotes the sign function; $k=|\bk|$)
\begin{equation*}
k_z= {\rm sgn}(\omega) \sqrt{\omega^2/c^2 - k^2} .
\end{equation*}
Similar expressions in terms of unit vectors
$\mbox{\boldmath\({\hat \epsilon}\)}^-_{\scriptstyle\rm TE}$ and $\mbox{\boldmath\({\hat \epsilon}\)}^-_{\scriptstyle\rm TM}$
hold for the field \mbox{${\bf E}^{\downarrow}({\bf k}, z, \omega) $} propagating towards the surface,
with the replacement $k_z \rightarrow - k_z$. The complete wave vector is denoted as ${\bf K}^\pm= {\bf k} \pm k_z \mathbf{\hat z}$,  and
the unit vectors given by (the frequency dependence will be generally omitted)
\begin{align*}
\mbox{\boldmath\({\hat \epsilon}\)}^+_{\scriptstyle\rm TE}({\bf k})&=\mbox{\boldmath\({\hat \epsilon}\)}^-_{\scriptstyle\rm TE}({\bf k})= \mathbf{\hat z} \times \mathbf{\hat k},
\\
\mbox{\boldmath\({\hat \epsilon}\)}^\pm_{\scriptstyle\rm TM}({\bf k})&= \mbox{\boldmath\({\hat \epsilon}\)}^\pm_{\scriptstyle\rm TE}({\bf k}) \times \mathbf{\hat K}^\pm ,
\end{align*}
correspond to transverse electric (TE) and  magnetic (TM) polarizations, respectively.

\begin{figure}[h]
\centering
\includegraphics[height=4cm]{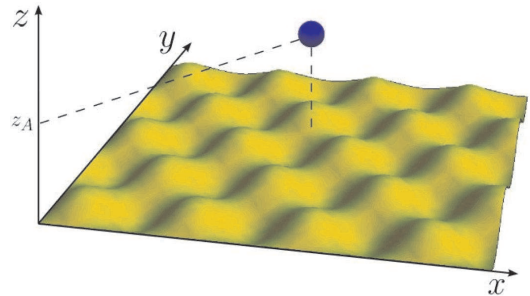}
\caption{ (color online)
Ground-state atom on top of a non-planar surface (profile function $h(x,y)$).
}
\label{pattern}
\end{figure}

The reflection by the non-planar surface modifies the wave vector $\bf k$
and the polarization $p={\rm TE,TM}$, while conserving the
frequency $\omega$  since the surface is assumed to be  stationary
\begin{equation}\label{RS1}
{E}^{\uparrow}_p({\bf k},\omega) = \int \frac{d^2{\bf k}'}{(2 \pi)^2}
\sum_{p'} \langle {\bf k},p | {\cal R}_S(\omega) | {\bf k}', p' \rangle \;
{E}^{\downarrow}_{p'}({\bf k}',\omega) .
\end{equation}
The reflection operator ${\cal R}_S(\omega)$ depends on the surface profile function $h(x,y)$.
Explicit results for its matrix elements are presented in Sec. 3 to first order in $h(x,y)$.

Let us assume for a moment that the $z$-axis is taken along the atom position ($ \br_A={\bf 0}$).
At zero temperature, the Casimir atom-surface interaction energy is then given by the zero-temperature
scattering formula as an integral over the positive imaginary frequency axis
($\omega\rightarrow i\xi$) \cite{Lambrecht06}
\begin{eqnarray}\label{scattering}
&&U({\bf R}_A)=\hbar\,\int_0^{\infty} \frac{d\xi}{2\pi}{\rm Tr}
\log\left(1-{\cal R}_S\,\,e^{- {\cal K} z_A}
\,{\cal R}_A\,e^{- {\cal K} z_A} \right), \nonumber\\
&&{\cal K}={\rm diag}(\kappa)  \;,\quad\kappa = \sqrt{\xi^2/c^2+k^2}
\end{eqnarray}
with ${\cal R}_A$ and ${\cal R}_S$ representing the reflection operators for the atom
and the surface, respectively. ${\cal K}$ is a diagonal operator in the basis of plane waves
$|\bk, p\rangle $ with eigenvalues $\kappa$,
so that $e^{- {\cal K} z_A}$ represents the propagation between the two scatterers.
Alternatively, it may also be interpreted as the displacement operator \cite{Emig07}:
here ${\cal R}_A$ is computed for a coordinate axis centered at the atom center-of-mass,
whereas $e^{- {\cal K} z_A}\,{\cal R}_A\,e^{- {\cal K} z_A}$ corresponds to the  `laboratory' axis.
A similar expression holds at finite temperature, replacing the integral over imaginary frequencies
$\xi$ by a sum over Matsubara frequencies.

It is actually simple to calculate the atomic reflection operator for an arbitrary position
$(\br_A, z_A)$ when taking the plane wave basis. The spherically-symmetric ground-state atom is described
as an induced electric dipole, with a dipole moment
\begin{equation}\label{dipole}
{\bf d}(\omega) = \alpha(\omega) {\bf E}({\bf R}_A,\omega).
\end{equation}
Let us emphasize at this point that $\alpha(\omega)$ is the dynamic polarizability defined
according to SI unit system \cite{esucgs}.
The dipole field components in the region $z<z_A$ are written in the representation
defined by \eqref{representation} with \cite{Carminati98}
\begin{equation}\label{dipoleC}
{E}^{\downarrow}_{({\rm dip}) p}({\bf k}, \omega) =
\frac{i \omega^2}{2\epsilon_0 c^2 k_z} \, \mbox{\boldmath\({\hat \epsilon}\)}
^-_{p}(\mathbf{k})\cdot {\bf d}(\omega) e^{-i\bk\cdot\br_A}e^{ik_z z_A}.
\end{equation}
When replacing \eqref{dipole} into \eqref{dipoleC}, we find two separate contributions associated
to the field Fourier components propagating upwards or downwards.
To calculate the atom reflection operator, we take the upward components
\begin{eqnarray}
\nonumber
&& {E}^{\downarrow}_{({\rm dip}) p}({\bf k}, \omega) =
\frac{i \omega^2\,\alpha(\omega)}{2\epsilon_0 c^2 k_z} \,
 \int \frac{d^2{\bf k}'}{(2 \pi)^2}
\sum_{p'}
\mbox{\boldmath\({\hat \epsilon}\)}
^-_{p}(\bk)\cdot \mbox{\boldmath\({\hat \epsilon}\)}
^+_{p'}(\bk') \\
\label{dipole_final}
&& \times
e^{-i(\bk-\bk')\cdot\br_A}e^{i(k_z + k_z')z_A}
{E}^{\uparrow}_{p'}({\bf k}', \omega) .
\end{eqnarray}
This result can be cast into a form similar to \eqref{RS1}, with the atomic reflection operator given by
 ($\omega\rightarrow i\xi$ and $k_z\rightarrow i\kappa$)
\begin{eqnarray}
\nonumber
 \langle \bk, p | {\cal  R}_A | \bk',p'\rangle  & = & -\frac{{\xi}^2}{2\kappa}\frac{\alpha(i\xi)}{\epsilon_0c^2}
 \hat{\epsilon}_{p}^{-}(\bk)\cdot \hat{\epsilon}_{p'}^{+}(\bk')
 \\
 \label{op_res_dip}
  & & \times  e^{-i(\bk-\bk')\cdot\br_A}e^{ - (\kappa + \kappa') z_A} .
\end{eqnarray}
The dependence on $\br_A$ and $z_A$ in this equation is exactly the one expected from the matrix elements
of the displacement operator in the plane wave basis. The displacement along the $z$-axis was already explicitly
taken into account in the scattering formula \eqref{scattering}, so we must
take $z_A=0$ when replacing \eqref{op_res_dip} into \eqref{scattering}.

As we assume that the atom-surface separation distance is much larger than the atomic dimensions,
we may expand the general scattering formula \eqref{scattering} to first order in $\alpha(\omega)$
or, equivalently, to first order in ${\cal R}_A$  \cite{foot_linear-alpha}
\begin{eqnarray}
\nonumber
U({\bf R}_A)& = &-\hbar   \int_0^{\infty} \frac{d\xi}{2 \pi}
\int \frac{d^2{\bf k}}{(2 \pi)^2}  \int \frac{d^2{\bf k}'}{(2 \pi)^2}
e^{ - (\kappa + \kappa') z_A}  \\
& \times &
\sum_{p,p'}
\langle {\bf k},p | {\cal R}_S | {\bf k}', p' \rangle
\langle {\bf k}',p' | {\cal R}_A | {\bf k}, p \rangle .
\label{vdW1}
\end{eqnarray}
The $z_A$ dependence is already explicit in \eqref{vdW1}  because  ${\cal R}_A$
in \eqref{scattering} was defined with respect to reference axis such that $z_A=0.$
Note that the resulting  $z_A$ dependence coincides with \eqref{op_res_dip},
which was derived for an arbitrary position with respect to the origin.
{}From \eqref{op_res_dip} and \eqref{vdW1}, we find
\begin{eqnarray}
&& U({\bf R}_A)  =  \frac{\hbar}{\epsilon_0c^2}\int_0^\infty \frac{d\xi}{2 \pi}  \xi^2\alpha(i \xi)
\int \frac{d^2{\bf k}}{(2 \pi)^2}  \int \frac{d^2{\bf k}'}{(2 \pi)^2}
\label{vdW2} \nonumber \\
&& \times
\frac{ e^{i ({\bf k} - {\bf k}') \cdot {\bf r}_A} e^{-(\kappa + \kappa') z_A}}{2 \kappa'} \\
&& \times
\sum_{p,p'}
\langle {\bf k},p | {\cal R}_S | {\bf k}', p' \rangle \;
\mbox{\boldmath\({\hat \epsilon}\)}^{+}_p({\bf k}) \cdot \mbox{\boldmath\({\hat \epsilon}\)}^{-}_{p'}({\bf k}') .
\nonumber
\end{eqnarray}
Note that this general formula holds for  magneto-dielectric  media, including the anisotropic case \cite{Rosa08}.
The dispersive potential for an atom with magnetic polarizability \cite{Feinberg} can also be derived
from \eqref{scattering} along the lines presented above. The
corresponding atomic reflection operator  is calculated in Appendix A.

As a first application of  \eqref{vdW2}, we briefly consider the case of a planar surface at $z=0$
(corresponding to a profile function $h(x,y)=0$) made of some isotropic material.
In this case, ${\cal R}_S$ is diagonal, its matrix elements being
\begin{equation}\label{Rzero}
\langle {\bf k}, p|{\cal R}_S | {\bf k}',p'\rangle = (2 \pi)^2 \delta^{(2)}({\bf k}-{\bf k}') \delta_{p,p'} r^p(k,i\xi),
\end{equation}
where $r^p(k,i\xi)$ are the specular reflection coefficients for the plane surface. For instance, for a homogeneous
non-magnetic bulk medium, they are given by the Fresnel formulas ($\epsilon(i\xi)=$ electric permittivity)
\begin{eqnarray}
&&r^{\rm TE}({\bf k},i\xi) = \frac{\kappa-\kappa_t}{\kappa+\kappa_t} \,,\;
r^{\rm TM}({\bf k},i\xi) = \frac{\epsilon(i\xi) \kappa -\kappa_t}{\epsilon(i\xi) \kappa+\kappa_t} ,
\label{Fresnel}\nonumber\\
&&\kappa_t=\sqrt{k^2+\epsilon(i\xi) \xi^2/c^2}
\end{eqnarray}
Eq.~\eqref{vdW2} then yields the known interaction energy  between a ground-state
atom and a plane surface valid for arbitrary separation distances  \cite{Barton, WylieSipe, Mostepanenko, DLP, Buhman05}
\begin{eqnarray}\label{specular}
U^{(0)}(z_A)   &  =  &  \frac{\hbar}{\epsilon_0c^2}\,\int_0^{\infty}
 \frac{d\xi}{2\pi}\, \xi^2\alpha(i \xi)\int  \frac{d^2\mathbf{k}}{(2\pi)^2}\,
\frac{e^{-2\kappa z_A}}{2 \kappa} \\
\nonumber
  & &   \times \left[ r^{\TE}(\bk,i\xi) - \Bigl(1+\frac{2c^2k^2}{\xi^2}\Bigr)r^{\TM}(\bk,i\xi) \right].
\end{eqnarray}
This expression can be simplified under the assumption of small or large distances $z_A$
(with respect to some typical atomic transition wavelength $\lambda_A$).
{}For $z_A\ll\lambda_A$ and $z_A\gg\lambda_A$, we get the van der Waals and Casimir-Polder limits
which correspond respectively to Eqs. (24) and (22) in Ref.~\cite{Antezza04}.

Eq.~\eqref{vdW2} represents a general result for the dispersive interaction between an atom and a body scatterer. The difficulty, of course, lies in the explicit derivation of the matrix elements of the corresponding operator ${\cal R}_S$. In the next section, we apply this result to derive the lateral dispersive  force for an atom on top of a corrugated surface up to first order in the corrugation profile $h(\br)$. A second interesting application, left for a future publication, would be to compute the roughness correction to the normal dispersive force -- in this case it is necessary to compute ${\cal R}_S$ up to second order in $h(\br)$ (see \cite{MaiaNeto05}).


\section{Perturbative interaction potential within the scattering approach}

In order to compute the exact atom-surface interaction potential it is necessary
to find the reflection operators of the surface, which is a highly non-trivial problem.
{}For the sake of illustration of the scattering formalism applied to the atom-surface
problem, we now compute those reflection operators in a perturbative expansion in terms
of the deviations $h({\bf r})$ of the surface profile with respect to the planar configuration.
We will calculate  the atom-surface interaction energy to first order in $h$.
This term gives rise both to a correction to the normal force and to a lateral force on the atom which does not exist in the case of a planar surface. We start by expanding the reflection operators in powers of $h$ as ${\cal R}={\cal R}^{(0)} + {\cal R}^{(1)} + O(h^2)$.
We model the optical response of
the homogeneous isotropic material
 in terms of its frequency-dependent electric permittivity $\epsilon(i\xi)$.
The zeroth-order term ${\cal R}^{(0)}$ is given by \eqref{Rzero}, whereas the first-order reflection matrix elements can be written as
\begin{equation}
\langle {\bf k}, p| {\cal R}^{(1)} | {\bf k}',p'\rangle =
R^{(1)}_{pp'}({\bf k},{\bf k}') H({\bf k}-{\bf k}') ,
\end{equation}
where $H({\bf k})$ is the Fourier transform of the surface profile $h({\bf r})$. Therefore, the first-order atom-surface interaction potential is
\begin{equation}
U^{(1)}({\bf R}_A) = \int \frac{d^2{\bf k}}{(2\pi)^2} e^{i {\bf k} \cdot {\bf r}_A} g(\bk,z_A) H({\bf k}) ,
\label{firstorder_potential}
\end{equation}
where $g(\bk,z_A)$ is the response function
\begin{eqnarray}
g(\bk,z_A) &&= \frac{\hbar}{\epsilon_0c^2} \int_0^{\infty} \frac{d\xi}{2\pi}
\xi^2\alpha(i\xi)\int \frac{d^2{\bf k}'}{(2\pi)^2} a_{{\bf k}',{\bf k}'-{\bf k}}
\label{g-def} \nonumber \\
a_{{\bf k}',{\bf k}''} &&= \frac{e^{-(\kappa'+\kappa'') z_A}}{2\kappa''} \\
&&\times\sum_{p',p''}
\hat{\bbm[\epsilon]}^{+}_{p'}({\bf k}') \cdot \hat{\bbm[\epsilon]}^{-}_{p''}({\bf k}'')R^{(1)}_{p'p''}({\bf k}', {\bf k}'').
\nonumber
\end{eqnarray}

The first-order reflection matrix was calculated in \cite{MaiaNeto05}
by following the approach presented in \cite{Greffet88} (see also \cite{Agarwal}).
It is a non-diagonal matrix with non-specular reflection coefficients given by
\begin{equation}
R^{(1)}_{pp'}({\bf k}, {\bf k}') = u_{pp'}(k,k';i\xi)\,
\Lambda_{pp'}^{(1)}({\bf k}, {\bf k}'; i\xi),
\end{equation}
where we have defined
\begin{equation}u_{pp'}(k,k';i\xi)=\frac{r^p(k,i\xi) t^{p'}(k',i\xi)}{t^p(k,i\xi)}.
\end{equation}
 $t^p({\bf k},i\xi)$ are the transmission Fresnel coefficients for
the planar interface defined by
\begin{eqnarray}
t^{\rm TE}({\bf k},i\xi) = \frac{2 \kappa}{\kappa+\kappa_t} &,&
t^{\rm TM}({\bf k},i\xi) = \frac{2 \sqrt{\epsilon(i\xi)} \kappa}{\epsilon(i\xi) \kappa+\kappa_t} .
\end{eqnarray}
The matrix ${\bf \Lambda}^{(1)}$ is expressed as ${\bf \Lambda}^{(1)}={\bf \Lambda}^{(1)}_+ - {\bf \Lambda}^{(1)}_-$
with
\begin{eqnarray}
&&{\bf \Lambda}^{(1)}_{\pm}({\bf k}, {\bf k}'; i\xi) =
(\kappa_t \pm \kappa) {\bf B}^{-1}_t
\left(
\begin{array}{cc}
C & S \\
- \frac{S}{1\pm \beta \beta_t} &  \frac{C \pm \beta \beta'_t}{1 \pm \beta \beta_t}
\end{array}
\right)
{\bf B}'_t ,
\nonumber\\
&& {\bf B}_t={\rm diag}(1,c \kappa_t/\sqrt{\epsilon(i\xi)} \xi), \quad
\beta=k/\kappa, \quad\beta_t=k/\kappa_t, \nonumber\\
&&C=\cos(\phi-\phi'),\quad S=\sin(\phi-\phi').
\end{eqnarray}
We have denoted with $\phi$ ($\phi'$) the angle between $\mathbf{k}$ ($\mathbf{k}'$) and an arbitrarily chosen axis on the surface plane.
Starting from these definitions and using the scalar product of polarization vectors
\begin{eqnarray}
\hat{\bbm[\epsilon]}^{+}_{\rm TE}({\bf k}) \cdot \hat{\bbm[\epsilon]}^{-}_{\rm TE}({\bf k}') &=& C , \nonumber \\
\hat{\bbm[\epsilon]}^{+}_{\rm TE}({\bf k}) \cdot \hat{\bbm[\epsilon]}^{-}_{\rm TM}({\bf k}') &=& \frac{c \kappa' S}{\xi} , \nonumber \\
\hat{\bbm[\epsilon]}^{+}_{\rm TM}({\bf k}) \cdot \hat{\bbm[\epsilon]}^{-}_{\rm TE}({\bf k}') &=& \frac{c \kappa S}{\xi} , \nonumber \\
\hat{\bbm[\epsilon]}^{+}_{\rm TM}({\bf k}) \cdot \hat{\bbm[\epsilon]}^{-}_{\rm TM}({\bf k}') &=& - \frac{c^2}{\xi^2} (k k' + \kappa \kappa' C) ,
\end{eqnarray}
we obtain for $a_{{\bf k}', {\bf k}''}$ the following general expression
\begin{widetext}
\begin{eqnarray}
&& a_{{\bf k}',{\bf k}''} = e^{-(\kappa'+\kappa'') z_A} \frac{\kappa'}{\kappa''}
\left[ C^2 u_{\rm TE,TE} + \frac{\kappa'' \kappa''_t S^2}{\sqrt{\epsilon(i\xi)} \xi^2/c^2} u_{\rm TE,TM}
\right. \nonumber \\
&& \left. + \frac{\sqrt{\epsilon(i\xi)} \kappa' \kappa'_t S^2}{\xi^2/c^2 - (\kappa')^2[\epsilon(i\xi) +1]}   u_{\rm TM,TE} +
\frac{ [k' k'' + \kappa' \kappa'' C][\epsilon(i\xi) k' k'' + \kappa'_t \kappa''_t C]}{(\xi^2/c^2) \{ \xi^2/c^2 - (\kappa')^2 [\epsilon(i\xi) + 1]  \} } u_{\rm TM, TM} \right].
\label{a-exact}
\end{eqnarray}
\end{widetext}
By inspection of Eq.~\eqref{a-exact}, one shows that the response function $g(\bk,z_A)$ depends only on the modulus $k=|{\bf k}|$
of the corrugation wave vector, as expected:
the information about the direction associated to the surface profile is
contained only in $H({\bf k})$, whereas the response function derived within the perturbative approach
is based on the planar geometry and its underlying  symmetry.

The results presented above allow for the numerical calculation of the
first-order potential for arbitrary ground-state
atoms and material media  by plugging
the corresponding polarizability and permittivity
functions into Eqs. (\ref{g-def}) and  (\ref{a-exact}).
 As an illustration, we consider a
 ground-state  rubidium atom above
a gold or a silicon surface. The dynamical atomic polarizability  is obtained from \cite{Babb}, and optical
data for gold and silicon are obtained from \cite{Palik}.
The values along the imaginary frequency axis are then calculated with the help of dispersion
relations \cite{Lambrecht00,Lambrecht04}.

In Fig.~2, we plot the response function after normalizing it by
\begin{eqnarray}
\label{CasimirPolder}
&&F_{\rm CP}^{(0)}(z_A)=-\frac{3\hbar c\alpha(0)}{8\pi^2\epsilon_0\,z_A^5},
\end{eqnarray}
that is also the result derived by Casimir and Polder \cite{CasimirPolder}
 for the attractive force on an atom on top
of a perfectly-reflecting plate at large separation distances.
The horizontal axis represents the  atom-surface separation $z_A$.
We take two different values of $k,$ corresponding to
corrugation wavelengths $\lambda=2\pi/k=10\,\mu{\rm m}$ and $100\,{\rm nm}.$

The global shape of the curves can be discussed in a qualitative manner,
which will be made more quantitative later on.
At low values of $kz_A$, PFA is expected to hold so that the response is approximately independent of $k.$
The ratio $g(k,z_A)/F_{\rm CP}$ thus grows linearly with $z,$
showing the well-known power-law change when sweeping from
the  unretarded van der Waals to the Casimir-Polder regime.
As $z_A$ approaches $1/k,$ PFA becomes gradually worse, and then a nontrivial
geometry effect reduces $g(k,z_A)$ exponentially  for $z_A\gg 1/k.$
In the next section, we show that geometry and real material effects
can be disentangled in most cases of interest,
thus providing a simple manner to explain how they jointly
produce the results shown in Fig.~2.

\begin{figure}[h]
\centering
\includegraphics[height=6.5cm]{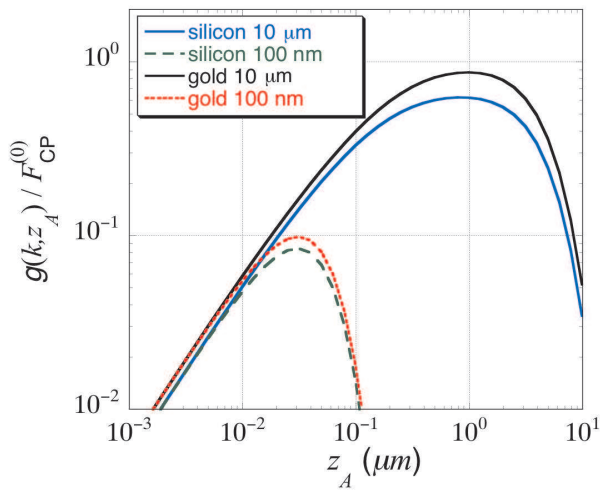}
\caption{ (color online) Dimensionless ratio
$g(k,z_A)/F_{\rm CP}^{(0)}(z_A)$
as a function of separation distance $z_A$ (for a ground-state Rb atom).
{}For the corrugation wavelength
$\lambda=2\pi/k=10\,\mu{\rm m},$
we show the values for gold (black) and silicon (blue) surfaces. We also show the results for
$\lambda=100\,{\rm nm}$ with gold (dotted red)
and silicon (dashed green) surfaces.
}
\label{pattern}
\end{figure}


\section{Disentangling geometry and real material effects}

As discussed in the Introduction, the most commonly used approximation methods to
compute atom-surface interactions are the proximity force approximation (PFA)
and pairwise summation approach (PWS).
In \cite{JPA} we compared these approximate methods with our exact scattering approach
to first-order in the surface profile for the case of perfect reflectors, and showed
that PFA overestimates and PWS underestimates the lateral atom-surface force. In this
section we want to expand upon these considerations and consider in further detail
the corrections to the PFA approximation beyond the case of perfect reflection.
This will allow us to disentangle the effects of geometry and
real materials, thus leading  to a very simple rule for the evaluation of $g(k,z_A).$

In the atom-surface context, the proximity force approximation amounts to compute
the atom-surface potential $U$ for a given geometry from the potential
for the planar geometry $U^{(0)}$ taken at the local atom-surface distance
\begin{eqnarray}\label{PFAbasico}
&&U\simeq U^{(0)}(z_A-h({\bf r}_A)) \simeq U^{(0)}-h({\bf r}_A) \partial_z U^{(0)}
\nonumber .
\end{eqnarray}
The PFA holds when the surface is very smooth in the scale of the separation distance,
that is in the limit $k z_A \rightarrow 0$. From our expression for the first-order
potential \eqref{firstorder_potential} one can indeed prove that for any material
the response function satisfies the ``proximity force theorem"
\begin{equation}\label{pft}
g(k=0,z_A) = - \partial_z U^{(0)}(z_A) .
\end{equation}
This follows from the fact that the response function at zero transverse momentum is given
by the specular limit of the non-specular coefficients $a_{{\bf k}',{\bf k}''}$,
which corresponds to ${\bf k}'={\bf k}''$. In this case we have
$R^{(1)}_{p',p''}({\bf k}',{\bf k}') = 2 \kappa' r^{p'}({\bf k}';i\xi) \delta_{p',p''}$,
and $a_{{\bf k}',{\bf k}'}=e^{-2 \kappa z_A} \sum_{p'}
\hat{\bbm[\epsilon]}^+_{p'}({\bf k}') \cdot \hat{\bbm[\epsilon]}^-_{p'}({\bf k}')
r^{p'}({\bf k}';i\xi)$. From (\ref{g-def}), the zero-momentum response function is then
\begin{eqnarray}
&&
g(k=0,z_A) = \frac{\hbar}{\epsilon_0c^2} \int_0^{\infty} \frac{d\xi}{2\pi} \xi^2
\alpha(i \xi) \nonumber \\
&& \times \int\frac{d^2 {\bf k}}{(2\pi)^2} e^{-2 \kappa z_A}
\sum_p
\hat{\bbm[\epsilon]}^+_{p}({\bf k}) \cdot \hat{\bbm[\epsilon]}^-_{p}({\bf k})
r^{p}({\bf k};i\xi).
\end{eqnarray}
Comparing with expression \eqref{specular} for $U^{(0)}$,
we immediately prove the proximity force theorem (\ref{pft}).

As a consequence of this discussion, we recover (\ref{PFAbasico}) from our more general
result (\ref{firstorder_potential}) when replacing $g(k,z_A)$ by $g(0,z_A).$
Now the value of $g(k,z_A)$ differs from $g(0,z_A)$ for any finite value of $k$
and it is worth quantifying deviations from PFA by introducing the function
\begin{equation}\label{rho}
\rho(k,z_A) = \frac{ g(k, z_A) }{ g(0,z_A) }.
\end{equation}
Deviations from PFA can first be discussed in the vicinity of $k=0$ where PFA is recovered
($\rho\rightarrow1$ for $k\rightarrow0$).
It is possible to deduce from \eqref{g-def}-\eqref{a-exact} that the first order derivative
of $\rho$ with respect to $k$ is identically zero, for any atomic polarizability and surface permittivity.
It follows that deviations from PFA  appear only at order $k^2$ in a Taylor expansion around  $k=0$
\begin{equation}\label{exp-k}
\rho(k,z_A) = 1+\frac{k^2}{2}\partial_k^2\rho(0,z_A)+O(k^3).
\end{equation}
{}For perfect reflectors, we can also prove that the second-order derivative
in (\ref{exp-k}) is negative, so that $\rho(k,z_A) $ is concave near $k=0.$

When going far beyond PFA (arbitrary values of $kz_A$), we have to compute $\rho(k,z_A)$ numerically.
The result of this calculation is plotted on Fig.~3, with $\rho$ shown as a function of $k z_A$
for three different values of  $z_A$ in the case of an Rb atom on top of a Si surface.
We notice that these curves are quite close to one another.
In order to obtain an estimate of their values, we also show the curve corresponding to the Casimir-Polder limit for perfect reflectors, which is derived from (\ref{rho})
and the calculations of Appendix B
\begin{equation}\label{rhoCP}
\rho_{\rm CP}^{\rm perf}(k,z_A)= e^{-kz_A} \left(1+ kz_A+ \frac{16 (kz_A)^2}{45} + \frac{(kz_A)^3}{45}\right).
\end{equation}
{}For this limiting case, $\rho$ depends only on the dimensionless ${\cal Z}=kz_A$.
We may thus conclude that this variable captures most of the geometry correction,
since $\rho$ depends very little on $z_A$ for a given ${\cal Z}.$
{}For distances larger than $1\,\mu{\rm m}$, $\rho$ may be well approximated by the CP formula Eq.~(\ref{rhoCP}).
This statement holds for the case of silicon, but the results for gold surfaces (not shown)
are even closer to the CP curve shown in Fig. 3.

Of course, finite conductivity corrections are very important for the evaluation of
the force between an atom and a plane plate, and the full response function $g$
has to be written as
\begin{equation}\label{disentanglement}
g(k,z)=\rho(k,z)\,\eta_F\, F^{(0)}_{\rm CP}
\end{equation}
In this formula, $F^{(0)}_{\rm CP}$ is the Casimir-Polder force (\ref{CasimirPolder}),
$\eta_F= F^{(0)}/F^{(0)}_{\rm CP}$ represents the reduction of the dispersive
force due to material properties of atom and surface (calculated for a plane geometry),
and $\rho$ describes the effect of geometry.

In Fig.~4, we plot $\eta_F$ as a function of $z_A$ for a Rb atom on top of a silicon or gold surface.
As expected, the CP result for perfect reflectors  is recovered
at large distances in the case of gold, whereas for silicon we find a reduction of
$\eta_F\sim 2/3$ due to the finite value of the zero-frequency permittivity in this case.
These results coincide with those obtained in \cite{Antezza04}. For both gold
and silicon surfaces, the reduction  gets stronger as the separation distance is decreased,
as expected since shorter distances correspond to larger frequencies, for which the
optical responses of both atom and surface are smaller. For very short distances, $\eta_F$ is linear
($\eta_F\approx 5.8\, z (\mu{\rm m}) $ for gold and $5.1 \,z  (\mu{\rm m})$ for silicon),
in agreement with  the power law modification expected in the van der Waals unretarded regime.

\begin{figure}[h]
\centering
\includegraphics[height=6.5cm]{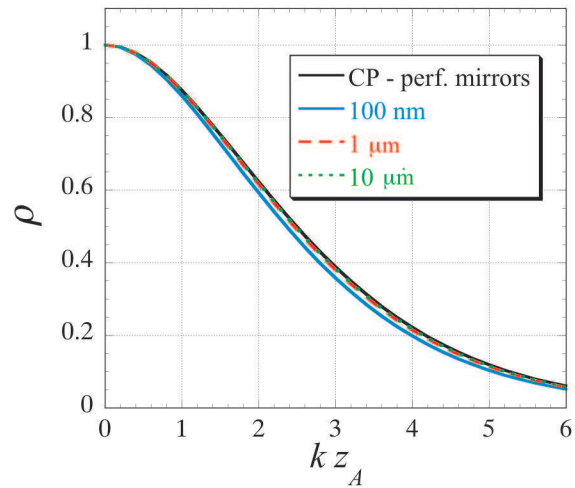}
\caption{ (color online)
Geometry correction factor $\rho$ as a function of $kz_A$ for a silicon surface
and $z_A=100\,{\rm nm}$ (light blue),
$1\,\mu{\rm m}$ (green), $10\,\mu{\rm m}$ (dashed black). The red line corresponds to the CP result
given by Eq.~(\ref{rhoCP}).
}
\label{pattern}
\end{figure}

Since $\rho$ is well approximated by the Casimir-Polder result for perfect reflectors (\ref{rhoCP}),
the real material  and beyond-PFA geometry corrections are approximately uncorrelated.
One may compute the former effect for the simple plane geometry ($\eta_F(z_A)$),
the latter for `perfect' atom  and surface ($ \rho^{\rm perf}_{\rm CP}(kz_A)$)
and finally combine the two effects with the help of (\ref{disentanglement}).
The simple analytical formula (\ref{rhoCP}) is thus of great practical relevance,
since it allows for an easy evaluation of the response function $g(k,z_A).$
{}For a given corrugation wavelength, as was discussed in connection with Fig.~2,
the two effects are more relevant at different separation ranges:
$\eta_F$ is mainly affected for small $z_A$
(producing the linear increase of all curves on the left part of Fig.~2),
whereas $\rho(k,z)$ differs for large $kz_A$
(leading to the exponential decay apparent on the right part of Fig.~2).
When $k$ is small enough (as in the example with $\lambda=10\,\mu{\rm m}$ on Fig.~2),
the two regions do not overlap, explaining why the maximum can approach unity.
Otherwise (as in the example with $\lambda=100\,{\rm nm}$ on Fig.~2), they do overlap
and the maximum value of the curve remains well below unity.

Only for very short separation distances  (or very short
corrugation wavelengths for a given $kz_A$) $\rho(k,z_A)$
 starts to deviate slightly from
$\rho^{\rm perf}_{\rm CP}(kz_A),$
as illustrated by the light blue curve in Fig.~3 for $z=100\,{\rm nm},$  resulting  in some entanglement between real material and geometry corrections.
In this case, $\rho$ is well approximated by the van der Waals analytical formula derived in Appendix C.

\begin{figure}[h]
\centering
\includegraphics[height=6.5cm]{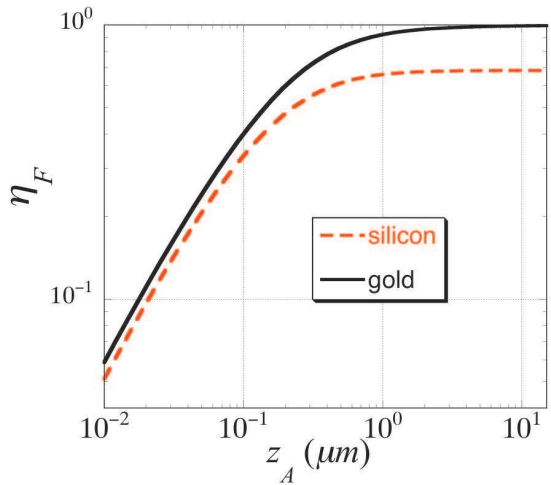}
\caption{ (color online)
Correction $\eta_F=F^{(0)}/F^{(0)}_{\rm CP}$ of the normal dispersive force between a
Rb ground state atom  and a plane gold (dark green) or silicon (red) surface as a function of the separation distance $z_A$.
}
\label{pattern}
\end{figure}


\section{Experimental prospects}

There are several possible experimental scenarios where non-trivial geometrical effects of atom-surface
forces could be probed. In particular, Bose-Einstein condensates represent an ideal sensor of quantum vacuum
forces since they are well controlled and characterized. One possible experiment is to use a BEC as a sensitive
oscillator, whose center-of-mass frequency is modified when the BEC is close to a surface \cite{Antezza05}.
This idea was recently used to measure the normal component of the Casimir-Polder force between a Rb condensate
oscillating along the perpendicular direction to a planar dielectric surface \cite{Cornell07}.
Lateral Casimir-Polder forces could be measured in a similar fashion with an elongated BEC oscillating parallel
to the surface, with it long axis parallel to the corrugation axis of a uni-axial corrugated surface \cite{JPA,PRL}.
We have shown in these references that non-trivial geometrical quantum vacuum effects, beyond the usual proximity
force and pair-wise summation approximations, could be measured using present-day cold atom technology.
We shall not further discuss this proposal here.

Rather, let us comment briefly on another possible scenario, that
although is less sensitive, has the nice feature of measuring directly the atom-surface interaction potential instead
of its second derivative, as in the BEC oscillator. The idea is to use a quasi one-dimensional BEC to locally probe
potential variations via imaging the local one-dimensional density. Small electric and magnetic fields of BECs in atom
chips have been recently probed using this technique \cite{Schmiedmayer}. Lateral Casimir-Polder forces could be measured
with a cigar-shaped BEC placed parallel to the surface and with its long axis perpendicular to the uni-axial corrugation
lines. The potential along the quasi one-dimensional BEC is related to the one-dimensional particle density $n_{1d}(x)$
as
\begin{equation}
V_{\rm ho}(x) + V_{\rm CP}(x) = -\hbar \omega_{\rm tr} \sqrt{1+4 a_{\rm scat} n_{1d}(x)},
\end{equation}
where $V_{\rm ho}(x)$ is the external harmonic trapping potential, $V_{\rm CP}(x)$ is the Casimir-Polder potential,
$\omega_{\rm tr}$ is the transverse (strong) trapping frequency, and $a_{\rm scat}$ is the s-wave scattering length.
In \cite{Schmiedmayer} the optimal potential single shot sensitivity $\Delta V$ is estimated as
$\Delta V= \gamma \Delta N / \rho_0^2 x_0$, where $\gamma=2\hbar^2 a_{\rm scat} /m$.
Here $m$ is the mass of Rb atoms, $\Delta N$ is the detection imaging noise, $x_0$ is the longitudinal spatial resolution,
and $\rho_0$ is the transverse spatial resolution. In the experiment \cite{Schmiedmayer}, $\omega_{\rm tr}=2 \pi \times 300$Hz,
$\Delta N \approx 4$ atoms per pixel in a CCD camera, leading to a single-shot single-point sensitivity to potential variations of
$\Delta V \approx 10^{-13}-10^{-14}$eV.

{}For the sake of estimating the order of magnitude of the lateral Casimir-Polder force
let us consider a uni-axial corrugated perfectly reflecting surface. We now use our perturbative expansion of the
potential in powers of the corrugation profile. In the retarded regime ($z_A \gg \lambda_A$),
the zeroth-order potential is $U^{(0)}(z_A)=-3 \hbar c \alpha(0) / 32 \pi^2 \epsilon_0 z_A^4$, where
$\alpha(0)/4\pi\epsilon_0=47.3 \times 10^{-30} {\rm m}^3$ for $^{87}$Rb atoms. For a
corrugation amplitude $h=100$nm, wavelength $\lambda=10\mu$m and an atom-plate distance $z_A=2\mu$m,
the correction is $U^{(1)}=1.13 \times 10^{-14}$eV, which is on the border of the experimental sensitivity
reported in \cite{Schmiedmayer}. The experimental
signal of the lateral Casimir-Polder force would consist on a density modulation of 1d BEC density following the law
$n_{\rm 1d}(x) \simeq [V_{\rm ho}(x)+ V_{\rm CP}(x)]^2$.

Other possible experiments to probe lateral Casimir-Polder forces could involve spin echo techniques for atomic beams flying
above corrugated surfaces \cite{DeKieviet2003}, or a two-component, phase-separated BEC "level" \cite{Timmermans2008}
used as a cold-atom analog of an AFM to map the corrugated surface potential.


\section{Conclusions}

We have developed a scattering approach to the dispersive force between
a ground state atom and a material body.
We have focused on the case of a corrugated surface, and derived
explicit results to first order in the corrugation amplitude.
Exact numerical calculation of the dispersive first-order potential
has been represented in terms of the response function $g(k,z_A),$
calculated for arbitrary values of the separation distance $z_A$ and
corrugation wavelength $\lambda=2\pi/k,$ as long as the corrugation
amplitude remains much smaller than both length scales.
Different atomic species and materials can be considered within our formalism.
Here we have illustrated our method by taking Rb atoms on top of
gold or silicon surfaces.

{}For  separation distances larger than $1\,\mu{\rm m},$ the response function
$g(k,z)$ can be easily obtained from the very simple CP result (\ref{rhoCP})
for the beyond-PFA geometry correction factor $\rho$,  together with
 the numerical values for the real material reduction factor
$\eta_F$ calculated for the plane geometry.

The position dependence of the resulting potential leads to a lateral dispersive force.
We have discussed a variety of possible experiments
with Bose-Einstein condensates that would allow for the verification of non-trivial
beyond-PFA effects with current state-of-the-art techniques.


\begin{acknowledgments}
We are grateful to James Babb for providing us with the
dynamic polarizability data for rubidium. R.M. acknowledges partial financial support by Ministero dell'Universit\`{a} e della Ricerca Scientifica e Tecnologica and by Comitato Regionale di Ricerche Nucleari e di Struttura della Materia. P.A.M.N. thanks  CNPq, CAPES and FAPERJ for financial support and ENS for a visiting professor position. A.L. acknowledges financial support from the French Carnot Institute LETI. D.A.R.D. acknowledges financial support from the U.S. Department of Energy through the LANL/LDRD Program. Part of this work was carried out at the Kavli Institute for Theoretical Physics, with support from NSF Grant No. PHY05-51164.
\end{acknowledgments}


\appendix

\section{Scattering matrix for an atomic magnetic dipole moment}

Eq.~\eqref{op_res_dip} gives the expression of the atomic reflection matrix elements
under the assumption that the atom is characterized by an electrical polarizability
$\alpha(\omega)$. The expression of the matrix elements of the operator $\mathcal{R}_A$
was found starting from the linear relation \eqref{dipole} between the induced atomic
electric dipole and the external (incoming) electric field calculated in the position
$\mathbf{R}_A$ and writing down the electric field produced by the induce dipole moment
$\mathbf{d}(\omega)$. In particular, we were interested in the field going towards the
surface $(\downarrow)$ as a function of the field coming from it $(\uparrow)$.

A completely analogous calculation can be performed assuming that the fluctuating magnetic
field $\mathbf{H}$ induces a magnetic dipole moment $\mathbf{m}$ on the atom given by
\begin{equation}\label{mag}
{\bf m}(\omega) = \beta(\omega) {\bf H}({\bf R}_A,\omega),
\end{equation}
where $\beta(\omega)$ is the magnetic polarizability. For the field generated by the
magnetic dipole in the region $z<z_A$ we find the Fourier component
\begin{eqnarray}
\nonumber
{E}^{\downarrow}_{({\rm mdip}) p}({\bf k}, \omega) & =&-
\frac{i\omega}{2\epsilon_0c^2k_z}e^{-i\bk\cdot\br_A}e^{ik_z z_A}\\
&&\times\mbox{\boldmath\({\hat \epsilon}\)}
^-_{p}(\mathbf{k})\cdot\bigl(\mathbf{K}^-\times{\bf m}(\omega)\bigr).\label{dipoleM}\end{eqnarray}
Using the following expression for the magnetic field coming upwards from the surface
\begin{equation}
{\bf H}^\uparrow(\mathbf{k},z,\omega)=\frac{1}{\mu_0\omega}\sum_p{E}_p^{\uparrow}({\bf k},
\omega)\bigl(\mathbf{K}^+\times\mbox{\boldmath\({\hat \epsilon}\)}
^+_p({\bf k},\omega)\bigr)e^{i k_zz},
\end{equation}
and replacing Eq.~\eqref{mag} into expression \eqref{dipoleM} we find
\begin{eqnarray}
\nonumber
&& \langle \bk, p | {\cal  R}^m_A | \bk',p'\rangle  = -\frac{\beta(i\xi)}{2\kappa}
e^{-i(\bk-\bk')\cdot\br_A}e^{ - (\kappa + \kappa') z_A} \nonumber \\
&& \times
\hat{\bbm[\epsilon]}_p^-(\mathbf{k},i\xi)\cdot\Bigl[\mathbf{K}^-\times\Bigl(\mathbf{K}'^+
\times\hat{\bbm[\epsilon]}_{p'}^+(\mathbf{k}',i\xi)\Bigr)\Bigr].
\label{Rmag}
\end{eqnarray}
As a consequence, if we want to include both contributions (electric and magnetic dipole moments),
the atomic reflection operator is simply given by the sum of the operators $\mathcal{R}_A$ and
$\mathcal{R}^M_A$, given by \eqref{op_res_dip} and \eqref{Rmag} respectively. It is possible to
check, for example, that for two atoms the scattering formula \eqref{scattering} gives back the
known result for the interatomic potential energy in the Casimir-Polder regime \cite{Feinberg}.


\section{ Casimir-Polder regime }

Simple closed forms for the first-order response function can be derived in some limiting cases.
In fact, $g(k,z_A)$ is given by the integral over the positive imaginary frequency axis
of a function involving three quantities that decay to zero: the atomic polarizability $\alpha(i\xi)$,
the round-trip propagation factor $\exp[-(\kappa'+\kappa'')z_A]$
and the reflection amplitudes $r^p(k,i\xi)$ (due to the frequency dependence of the permittivity function).
These  functions decay on different frequency scales:
$c/\lambda_A$, $c/z_A$ and $\omega_P$ (a typical frequency associated to the surface optical response,
for example the plasma frequency for metallic media), respectively.
As a consequence, the significant range of frequencies giving the main contribution to the integral
in (\ref{g-def}) will be determined by the smallest of these frequency scales. One can then expect
that the integral (\ref{g-def}) can be considerably simplified for some particular relations between
these frequency scales.

In this Appendix,
we  consider the case $z_A\gg\lambda_P,\lambda_A$, usually referred to as Casimir-Polder regime
(note that we assume $z_A \ll \hbar c/(k_B T)$ so as to neglect thermal corrections).
Since the smallest frequency scale is $c/z_A,$
we can replace the dynamical polarizability $\alpha(i\xi)$ and the permittivity
$\epsilon(i\xi)$ by their  zero frequency values. For the metallic case,
we can take $\epsilon(0)\rightarrow\infty$ and then derive remarkably
simple expressions from (\ref{a-exact}):
\begin{eqnarray}
&& a_{{\bf k}',{\bf k}''} = \frac{1}{2} e^{-(\kappa'+\kappa'') z_A}
\label{a-perfect-reflector}  \\
&& \times  \left\{\frac{1}{\kappa' \kappa''} [k^2 + (\kappa'-\kappa'')^2] + \frac{c^2}{\xi^2}
[k^2 - (\kappa'+\kappa'')^2]
\right\} .
\nonumber
\end{eqnarray}
This expression could also have been obtained by taking the usual perfectly-reflecting
boundary conditions on the corrugated surface.

After replacing (\ref{a-perfect-reflector}) into
\begin{equation}
g_{\rm CP}(k,z_A) = \frac{\hbar\alpha(0)}{\epsilon_0c^2}\int_0^{\infty}
\frac{d\xi}{2\pi} \xi^2 \int \frac{d^2{\bf k}'}{(2\pi)^2} a_{{\bf k}',{\bf k}''}(z_A,\xi) ,
\nonumber
\end{equation}
we change the integration variables from
$(\xi,k',\phi')$ to $(\kappa',\kappa'',\varphi)$, taking
$k'(\kappa',\kappa'',\varphi)=[(\kappa')^2-(\kappa'')^2+ k^2]/(2 k \cos \phi')$ and
$\xi(\kappa',\kappa'',\phi')=c [(\kappa')^2-(k')^2(\kappa',\kappa'',\phi')]^{1/2}$.
The Jacobian of the transformation is
$|J|=2 c\kappa' \kappa'' / \{4 k^2 (\kappa')^2 \cos^2\phi' - [(\kappa')^2-(\kappa'')^2 + k^2] \}^{1/2}$.
We first  use the property $\int_0^{2\pi}d\phi'(...)=2\int_0^\pi d\phi'(...)$. Then we
write the integral as a sum of two contributions:
\[
\int_0^{\infty} d\kappa' \int_{|\kappa'-k|}^{\sqrt{(\kappa')^2+k^2}} d\kappa'' \int_0^{\phi_m} d\phi'
\]
and
\[
\int_0^{\infty} d\kappa' \int_{\sqrt{(\kappa')^2+k^2}}^{\kappa'+k}d\kappa'' \int_{\pi-\phi_m}^\pi d\phi',
\]
with $\cos \varphi_m =[(\kappa')^2-(\kappa'')^2 + k^2]/2\kappa' k,$
and introduce further auxiliary integration variables $u=\kappa'+\kappa''$ and $v=\kappa'-\kappa''$
to obtain the final analytical result
\begin{eqnarray}
&&g_{\rm CP}^{\rm perf}(k,z_A)= F_{\rm CP}^{(0)}(z_A)e^{-\mathcal{Z}} \left(1+ \mathcal{Z} + \frac{16 \mathcal{Z} ^2}{45}
+ \frac{\mathcal{Z} ^3}{45}\right)
\nonumber\\
&& \mathcal{Z} \equiv kz_A
\end{eqnarray}


\section{Van der Waals regime}

In this appendix, we assume that $z_A \ll \lambda_A$.
In this  van der Waals regime, one can neglect retardation effects
and assume an instantaneous atom-surface interaction (limit $c\to\infty$).

In order to extract some simple expressions for the vdW response function,
we have first to choose a specific expression for the atomic polarizability $\alpha(i\xi)$.
We will use the following model function suitable for a multilevel isotropic atom having
transition frequencies $\omega_{n0}$ from the $n$-th excited state to the ground state and
electric dipole matrix elements $d_{n0}$ on these transitions
\begin{equation}
\alpha(i\xi)=\frac{2}{3\hbar}\sum_n\frac{\omega_{n0}d_{n0}^2}{\omega_{n0}^2+\xi^2},
\end{equation}
Then we have to specify the transmission and reflection amplitude functions.
Here we take Fresnel formulas with two
different models for the permittivity $\epsilon(i\xi)$.

{\it Plasma metals.} We first take the plasma model for metallic materials:
\begin{equation}
\epsilon(i\xi)=1+\frac{\omega_P^2}{\xi^2}.
\end{equation}
In this case we  obtain from (\ref{g-def})-(\ref{a-exact})
\begin{widetext}
\begin{equation}
\label{vdWgeral}
g_{\rm vdW}(k,z_A)=-\sum_n\frac{kd_{n0}^2x_n}{192\sqrt{2}\pi\epsilon_0z_A^3(x_n^2+2\sqrt{2}x_n+2)}
\Bigl[6\sqrt{2}{\cal Z}(x_n+\sqrt{2})K_0({\cal Z})+(\sqrt{2}({\cal Z}^2+12)x_n+{\cal Z}^2+24)K_1({\cal Z})\Bigr] ,
\end{equation}
\end{widetext}
where $x_n=\omega_P/\omega_{n0}$ and ${\cal Z}=kz_A$.
If we let all the $x_n$ go to infinity, we get the result for a perfect conductor \cite{PRL}
\begin{eqnarray}
g^{\rm perf}_{\rm vdW}(k,z_A) &=& -\sum_n\frac{kd_{n0}^2}{192\pi\epsilon_0z_A^3} \\
&& \times \Bigl[6{\cal Z}K_0({\cal Z})+({\cal Z}^2+12)K_1({\cal Z})\Bigr] . \nonumber
\end{eqnarray}
In the opposite limit, with $\omega_P\ll \omega_{n0}$ for all $n,$ we
derive from (\ref{vdWgeral}) the plasmon van der Waals  limit
\begin{eqnarray}
g_{\rm vdW}^{\rm plas}(k,z_A) &=& -\sum_n\frac{kd_{n0}^2x_n}{384\sqrt{2}\pi\epsilon_0z_A^3} \\
&& \times\Bigl[12{\cal Z}K_0({\cal Z})+({\cal Z}^2+24)K_1({\cal Z})\Bigr] . \nonumber
\end{eqnarray}
This result is, as expected, proportional to the plasma frequency $\omega_P$.

{\it Semiconductors:} In order to treat  the case of semiconductors, we assume that the material
can be described by the Drude-Lorentz model function
\begin{equation}\epsilon(i\xi)=1+\frac{\omega_{DL}^2(\epsilon(0)-1)}{\omega_{DL}^2+\xi^2}.\end{equation}
In this case we get
\begin{widetext}
\begin{equation}\begin{split}
g_{\rm vdW}(k,z_A)&=-\sum_n\frac{\gamma kd_{n0}^2x_n}{384\pi\epsilon_0z_A^3(\gamma+2)^\frac{3}{2}((\gamma+2)x_n^2-2)^2}
\Bigl(12A_0{\cal Z}K_0({\cal Z})+A_1({\cal Z})K_1({\cal Z})\Bigr)\\
A_0&=(\gamma+2)(\sqrt{\gamma+2}x_n-\sqrt{2})((\gamma+2)x_n^2-2)\\
A_1&=2(\gamma+2)^\frac{3}{2}(({\cal Z}^2+12)\gamma+24)x_n^3-3\sqrt{2}(\gamma+2)(({\cal Z}^2+8)\gamma+16)x_n^2-48(\gamma+2)^\frac{3}{2}x_n+2\sqrt{2}(({\cal Z}^2+24)\gamma+48)\\\end{split}\end{equation}
\end{widetext}
where $\gamma=\epsilon(0)-1$ and $x_n=\omega_{DL}/\omega_{n0}$.



\begin{thebibliography}{99}
\bibitem{esucgs}Dimensionally speaking, $\alpha(\omega)$ is equal to $4\pi\epsilon_0$ times a volume.
The latter volume (i.e. $\alpha(\omega)/4\pi\epsilon_0$) is identified with the polarizability
in the ``electrostatic cgs'' unit system,  used in a number of papers studying dispersion forces.
This point has to be treated carefully when comparing various expressions.
\bibitem{foot_linear-alpha} For consistency,  higher-order multipoles beyond the electric dipole
approximation for the atom-field coupling  would also be needed to compute
corrections involving higher powers of ${\cal R}_{A}$.
\end{thebibliography}
\end{document}